\documentclass[9pt, conference, letterpaper]{IEEEtran} 
\usepackage[paperwidth=8.5in, paperheight=11in,
            top=1in, left=0.75in,
            textwidth=7in, textheight=9in]{geometry}
            
\usepackage{times}
\usepackage{titling}
\usepackage{multicol}
\usepackage{ragged2e}
\justifying

\date{}           

\usepackage{caption}
\usepackage{cite}
\usepackage{amsmath,amssymb,amsfonts}
\usepackage{algorithmic}
\usepackage{graphicx}
\usepackage{textcomp}
\usepackage{xcolor}
\usepackage{relsize}

\makeatletter
\renewcommand{\section}{\@startsection{section}{1}{\z@}%
   {-2ex plus -0.5ex minus -0.2ex}%
   {1ex plus 0.3ex}%
   {\centering\normalfont\bfseries\normalsize\uppercase}}%

\renewcommand{\subsection}{\@startsection{subsection}{2}{\z@}%
  {-1.8ex plus -0.5ex minus -0.2ex}%
  {0.8ex plus 0.2ex}%
  {\normalfont\bfseries\normalsize}}%

\makeatother
\usepackage{etoolbox}
\makeatletter
\patchcmd{\@startsection}
  {\@afterindenttrue}
  {\@afterindentfalse}
  {}{}
\makeatother

\DeclareMathOperator{\diag}{diag}

\title{\fontsize{12pt}{16pt}\selectfont\bfseries\uppercase{Information Entropy-Based Scheduling for Communication-Efficient Decentralized Learning}\\}

\author{
  Jaiprakash Nagar\IEEEauthorrefmark{1}, 
  Zheng Chen\IEEEauthorrefmark{2}, 
  Marios Kountouris\IEEEauthorrefmark{1}\IEEEauthorrefmark{3}, 
  Photios A. Stavrou\IEEEauthorrefmark{1} \\  
  \IEEEauthorrefmark{1}Communication Systems Department, EURECOM, Sophia-Antipolis, France \\
  \IEEEauthorrefmark{2}Department of Electrical Engineering, Linköping University, Sweden \\
  \IEEEauthorrefmark{3}DaSCI, Department of Computer Science and Artificial Intelligence, University of Granada, Spain
}

\begin{document}
\maketitle

\thispagestyle{empty}
\begin{abstract}
This paper addresses decentralized stochastic gradient descent (D-SGD) over resource-constrained networks by introducing node-based and link-based scheduling strategies to enhance communication efficiency. In each iteration of the D-SGD algorithm, only a few disjoint subsets of nodes or links are randomly activated, subject to a given communication cost constraint. We propose a novel importance metric based on information entropy to determine node and link scheduling probabilities. 
We validate the effectiveness of our approach through extensive simulations, comparing it against state-of-the-art methods, including betweenness centrality (BC) for node scheduling and \textit{MATCHA} for link scheduling. The results show that our method consistently outperforms the BC-based method in the node scheduling case, achieving faster convergence with up to 60\% lower communication budgets. At higher communication budgets (above 60\%), our method maintains comparable or superior performance. In the link scheduling case, our method delivers results that are superior to or on par with those of \textit{MATCHA}.
\end{abstract}

\section{Introduction}
\label{sec:intro}
Collaborative training of a machine learning (ML) model with decentralized data and model aggregation offers benefits over centralized learning, such as lower communication costs, enhanced privacy, increased scalability, and greater robustness against a single point of failure\cite{hu2021distributed, beltran2023decentralized}. In such setups, nodes train a shared ML model on their local datasets beyond sharing it with other network nodes, thus preserving user privacy \cite{mcmahan2017communication}. Then, each node shares its locally trained model with the rest of the network nodes by peer-to-peer communication among neighboring nodes, known as decentralized learning. 

Decentralized stochastic gradient descent (D-SGD) is a widely adopted optimization approach for decentralized learning \cite{lian2017can}. In every iteration of D-SGD, each node computes its stochastic gradient, combines it with the updates received from the neighbors, and then shares the updated model with their neighbors. The frequent exchange of model updates results in significant communication overhead, making the communication cost a bottleneck. Previously, the convergence speed of D-SGD was analyzed in terms of the decrease in error per iteration \cite{wang2021cooperative, jakovetic2018convergence}, neglecting the communication cost of each iteration. However, communication dynamics significantly affect the convergence speed of D-SGD \cite{nedic2018network}, as the choice of the medium access scheme governs the communication cost per iteration. Based on this observation, some works, e.g., \cite{wang2019adaptive, wang2022matcha, chiu2023laplacian}, used graph sparsification to enhance convergence speed by reducing communication frequency, motivated by the idea that the link scheduling policies proposed in these works prioritize more important links, allowing them to be activated more frequently and thereby achieving faster convergence under limited communication cost. In addition, a node can transmit model updates to all its neighbors in a single transmission slot by exploiting the broadcast nature of communication networks. A more recent work in \cite{herrera2024faster} proposed a node scheduling policy that accounts for the broadcast effect and optimized the activation probabilities of collision-free node subsets.

In light of the existing literature, a challenging question is how to measure the importance of links and nodes in a network. In communication networks, each link contributes differently to connectivity—some have minimal impact, while others are critical. 
Similarly, the importance of a node in a network reflects its role and impact on the structure and dynamics of the network. Furthermore, computing the importance of the node helps assess the vulnerability of the network to node failures and the influence of the node on the flow of information within the network \cite{yin2019node}. In the context of decentralized learning, the importance of nodes should be considered when assigning scheduling probabilities to maximize the convergence speed. Several methods exist for computing link/node importance in a graph, e.g. \cite{yang2019node}, \cite{herrera2023distributed}, the work presented in \cite{herrera2024decentralized} considers the betweenness centrality (BC) metric as it renders more accurately the importance of nodes than other centrality measures. However, the BC method is less effective in identifying critical nodes in large and densely connected topologies. Moreover, BC is unsuitable for irregular topologies, as nodes with a degree of one always receive a BC value of zero, regardless of their potential importance.

This paper builds upon the work presented in \cite{herrera2024decentralized}, which employs a BC-based node scheduling approach for decentralized learning with partial communication. In contrast, we introduce a novel information entropy (IE)-based metric to assess the importance of links and nodes. This metric identifies more accurately critical components of the network than existing centrality measures. By quantifying the information richness and diversity of a node’s local connections, our approach directly captures its structural complexity and the influence of its immediate environment. As a result, node importance is evaluated in a manner that is more computationally efficient and robust, particularly in large, complex, and densely connected networks where local structure is more relevant than global shortest paths. Consequently, the proposed method overcomes the key limitations of traditional centrality-based approaches, enabling efficient and scalable computation of node and link importance in both sparse and dense, irregular topologies. 
In addition, the proposed approach avoids monotonic importance ranking, enabling more accurate learning with fewer transmission slots. Experimental results demonstrate that the proposed IE-based link importance metric consistently performs as well as, or better than, the optimized MATCHA framework. Moreover, the IE-based node importance metric significantly outperforms the BC-based approach, achieving higher test accuracy and lower training loss for the D-SGD algorithm while using fewer communication rounds. 

\begin{figure}[t]
    \centering
    \includegraphics[width=0.6\linewidth]{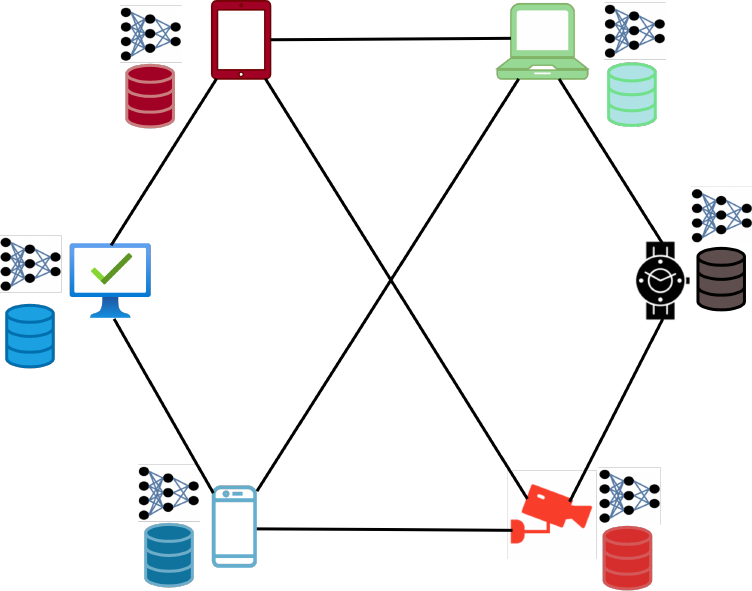}
    \captionsetup{justification=centering}
    \caption{\vspace{-1em}Distributed machine learning scenario.}
    \label{fig:dml}
\end{figure}

\paragraph*{\bf Notation:} A connected network with $N$ nodes is represented by an undirected graph $G = \left(V, E\right)$, referred to as the base topology. The node set is denoted by ${V} = \{1,\ldots, N\}$, and the edge set $E \subseteq \{(i,j) \in {V} \times {V}| i \neq j\}$ specifies which pair of nodes can directly communicate. For each node $i \in V$, its set of neighbors is defined as ${\cal N}_{i}=\{j|(i,j) \in E\}$, and its degree is given by $d_i=|{\cal N}_{i}|$. The topology of the graph $G$ can be described by its adjacency matrix $\mathbf{A} \in \mathbb{R}^{N \times N}$, where $A_{ij}=1$ if $(i,j) \in E$ and $A_{ij}=0$ otherwise. The Laplacian matrix $\mathbf{L}$ associated with the graph is defined as $\mathbf{L}:= \mathbf{D}-\mathbf{A}$, where $\mathbf{D}$ is a diagonal matrix with entries $D_{ii}=d_i$, representing the degree of node $i$.

\section{Decentralized Machine Learning} 
\label{sec:DCM}

We consider a network consisting of $N$ nodes, where each node $i \in V$ has access to a local training dataset ${\cal D}_{i}$. The primary objective is for all nodes to collaboratively train an ML model by sharing model parameters with their immediate neighbors (see Fig. \ref{fig:dml}). Specifically, each node aims to estimate a common parameter vector $\mathbf{x} \in \mathbb{R}^{d}$ that minimizes the following global objective function: $F(\mathbf{x}):=\frac{1}{N} \sum_{i=1}^N F_i(\mathbf{x})$, where $F_i(\mathbf{x})$ denotes the local objective function at node $i$, defined as $F_i(\mathbf{x}):=\frac{1}{|{\cal D}_i|} \sum_{s \in {\cal D}_i}  \ell (\mathbf{x};s)$, and $\ell (\mathbf{x};s)$ is the loss function evaluated in the sample $s$ from the local dataset ${\cal D}_{i}$.

We adopt a consensus-based D-SGD algorithm for collaborative model training,  where each node $i$ computes local gradients using its own dataset ${\cal D}_{i}$, and performs consensus-based model averaging with its neighbors \cite{herrera2023distributed}. Specifically, each iteration of the algorithm consists of the following steps:

\begin{itemize}
\item \textbf{Stochastic gradient computation:} At iteration $k$, node $i$ computes the gradient vector $\mathbf{g}_i^{(k)} = \frac{1}{|\zeta|} \sum_{s \in \zeta} \mathbf{\nabla} \ell ({\mathbf{x}_i}^{(k)};s)$, where $\zeta \subseteq {\cal D}_{i}$ is a randomly sampled mini-batch from the local dataset ${\cal D}_{i}$. The node then performs a local update of its model parameters using the stochastic gradient: $\mathbf{x}_i^{\left(k+\frac{1}{2}\right)}=\mathbf{x}_i^{(k)}-\gamma \mathbf{g}_i^{(k)}$, where $\gamma$ is the learning rate.

\item \textbf{Communication with neighbors:} Each node exchanges model parameters with its immediate neighbors to facilitate consensus. Specifically, node $i$ first transmits its updated model parameters $\mathbf{x}_i^{\left(k+\frac{1}{2}\right)}$ to its neighbors, and then receives the corresponding model parameters from them, denoted ${\cal N}_{i}$. As a result, node $i$ obtains both its own and its neighbors' intermediate model parameters for the current iteration. 

\item \textbf{Consensus-based model averaging:} Each node $i$ updates its model parameters by averaging its own and its neighbors' intermediate updates using $\mathbf{x}_i^{(k+1)}= \sum_{j=1}^N W_{i j}^{(k)} \mathbf{x}_j^{\left(k+\frac{1}{2}\right)}$, where $W_{i j}^{(k)}$ denotes the weight assigned by node $i$ to the model parameters received from node $j$ during iteration $k$. These weights can be collectively written in the form of a \textbf{\textit{weight}} or a \textbf{\textit{mixing matrix}} $\mathbf{W}^{(k)} \in \mathbb{R}^{N \times N}$. Note that $W_{i j} \neq 0$ only if $(i,j) \in E$, i.e., if node $j$ is a neighbor of node $i$.
\end{itemize}

\subsection{Mixing Matrix Properties and Design}
According to \cite{ye2022decentralized}, the convergence of the D-SGD algorithm is guaranteed under the following sufficient conditions:

\begin{itemize}
    \item the mixing matrix $\mathbf{W}$ is symmetric and doubly stochastic;
    \item the spectral gap of $\mathbf{W}$ is strictly positive.
\end{itemize}

In addition to the aforementioned conditions on $\mathbf{W}$, the convergence of the D-SGD algorithm also requires that the variances of the stochastic gradients $\mathbf{g}_i$ are bounded and that the local objective functions $F_i(\mathbf{x})$ are differentiable with Lipschitz-continuous gradients \cite{herrera2024faster}. A common and effective design for the mixing matrix is based on the graph Laplacian matrix \cite{wang2022matcha}, and is given by $\mathbf{W} := \mathbf{I} - \alpha \mathbf{L}$, where $\mathbf{I}$ is the identity matrix, $\alpha > 0$ is a tunable parameter, and $\mathbf{L}$ is the Laplacian matrix of the communication graph. The parameter $\alpha > 0$ should be chosen to ensure that the spectral radius satisfies $\rho(\mathbf{W} - \mathbf{J}) < 1$, where $\rho(\cdot)$ denotes the spectral radius of a matrix, $\mathbf{J} = \frac{1}{N}\mathbf{11^T}$, and $\mathbf{1}$ is the all-ones column vector. Furthermore, for faster convergence, $\alpha$ should be selected to minimize $\rho(\mathbf{W} - \mathbf{J})$.  

\section{Communication-Efficient Decentralized Learning under Partial Communication}
In this section, we describe the communication model, introduce the proposed scheduling schemes based on node importance metrics, and explore strategies for optimizing the mixing matrices. To ensure convergence in distributed learning, nodes must exchange model parameters with their neighbors according to the communication model. Due to the broadcast nature of wireless channels, a node can share its updated model with all neighbors in a single time slot. However, to prevent interference and packet collisions from simultaneous transmissions, nodes must coordinate their communication schedules with neighboring nodes, ensuring reliable communication and algorithmic convergence.

The most effective way to ensure collision-free communication is by allocating orthogonal communication resources, such as distinct time slots, to each broadcasting node. In this scheme, every node is assigned a dedicated time slot during which it can broadcast its model parameters to its neighbors without interference. Consequently, a complete communication round, in which all nodes exchange updates, requires a total of $N$ time slots to ensure full information fusion across the network. Thus, under a full communication scenario, executing $\mathcal{K}$ iterations of the consensus algorithm necessitates a total of $N\mathcal{K}$ time slots.

 \begin{figure}[t]
    \centering
    \includegraphics[width=1\linewidth]{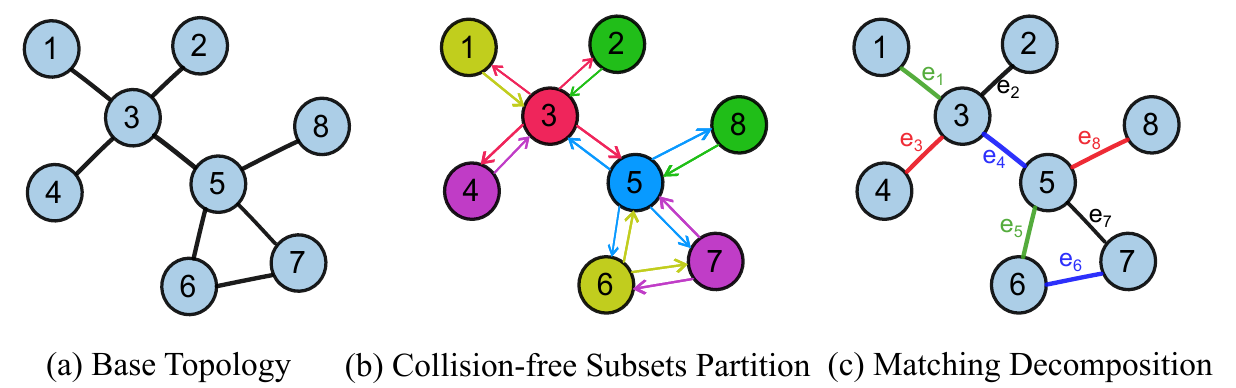}
    \captionsetup{justification=centering}
    \caption{Partitioning (a) Undirected base topology into (b) Collision-free Subsets and (c) Matchings.}
    \label{fig:base}
\end{figure}

\subsection{Communication Model}
We consider two types of communication scenarios: (1) peer-to-peer communication with link-based scheduling and (2) broadcast transmission with node-based scheduling. In the first scenario, links are activated randomly and each information exchange over a link requires two transmission slots. In the second, nodes are activated randomly and each broadcast allows a node’s model update to reach multiple neighbors simultaneously. In both cases, we assume that concurrent transmissions adhere to collision-free conditions.

\subsubsection{Link Scheduling with Peer-to-Peer Transmission}
A matching is a subgraph of $G$ where each node is connected to at most one edge. The base topology is decomposed into $M$ disjoint matching subgraphs $\{ G_j(V, E_j) \}_{j=1}^{M}$, where $E = \bigcup_{j=1}^{M} E_j$ and ${E}_{i} \cap { E}_{j} = \emptyset, \forall i \neq j$. Fig. \ref{fig:base}(c) shows the matching decomposition of a base topology using the edge-coloring algorithm \cite{misra1992constructive}, where edges with the same color form a matching. Note that in each communication round, only a subset of these matchings is activated according to the scheduling policy. This scheduling is guided by the principle that the critical links that maintain network connectivity should be activated more frequently than the less essential ones.

\subsubsection{Node Scheduling with Broadcast}
A collision-free subset is a group of nodes in which no two nodes share a common neighbor, allowing them to transmit model updates simultaneously without the risk of collisions. The base topology is partitioned into a finite number of such collision-free subsets, denoted by $q$. Fig. \ref{fig:base}(b) illustrates the partitioning of the base topology, where nodes of the same color form a collision-free subset that requires only a single transmission slot. Each communication round of the D-SGD algorithm is further divided into multiple transmission time slots. For a given node pair $(i,j)$, node $j$ can successfully receive information from node $i$ only if neither node $j$ nor any of its neighbors is transmitting during that slot.

An undirected graph $G = \left(V, E\right)$ can be represented by a directed graph $G^ {\textit {d}} = \left(V, E^ \textit{d}\right)$, where each undirected edge is replaced by a pair of bidirectional directed links connecting each pair of nodes. Due to the broadcast nature of communication channels, each node transmits its information to all of its neighbors via directed links, resulting in a local star topology. To enable collision-free concurrent transmissions, the base topology is partitioned into disjoint subsets of nodes $\{{\cal V}_{r}\}_{r \in [q]}$, where $q\leq N$. Each subset ${\cal V}_{r}$ contains nodes that do not share any common neighbors, allowing them to broadcast simultaneously without interference, $i, j \in {\cal V}_{r}$ if ${\cal N}_{i} \cap {\cal N}_{j} = \emptyset$. These subsets form a complete partition of the node set, satisfying $\bigcup_{r=1}^{q} {\cal V}_{r} = {\cal V}$, ${\cal V}_{r} \cap {\cal V}_{l} = \emptyset, \forall r \neq l$.

We employed a greedy vertex-coloring algorithm \cite{bollobas1998modern} to partition the base topology by assigning different colors to connected nodes. To enforce the constraint that no two nodes sharing a common neighbor can broadcast concurrently, even if they belong to the same subset, we construct an auxiliary graph $G^{\textit{a}} = \left(V, E^\textit{a}\right)$, where $E^\textit{a}$ includes all original edges from $E$ along with additional edges connecting every pair of nodes that share at least one common neighbor. This ensures that nodes with the same color, and thus in the same subset, form a collision-free set. The total number of such subsets, denoted by $q$, corresponds to the number of colors assigned by the coloring algorithm.

Furthermore, this work considers a partial communication policy aimed at reducing the number of transmission slots per iteration and minimizing the overall communication cost required to achieve consensus in a distributed learning algorithm. Under partial communication, only a subset of network links and nodes is activated for information exchange in each round (see, e.g., \cite{herrera2024decentralized}), implying that communication should not be uniformly distributed across all links and nodes. Instead, the frequency of communication should be guided by the relative importance of individual links and nodes within the graph. To this end, this paper employs an information entropy-based approach to quantify the importance of links and nodes, as detailed in Section~\ref{Node Importance}.

\subsection{Node Importance Vector}
\label{Node Importance}
In what follows, we describe the IE-based node importance approach, which leverages node degree to compute the self-information (SI) \cite{shannon:1948} of an edge connecting two adjacent nodes \cite{liu2022self}. Specifically, the method quantifies the informational contribution of a link $(i,j) \in E$ using the degrees of the two nodes it connects, $d_i$ and $d_j$. The probability associated with the existence of link $(i,j)$ is defined as $P(i,j) = \frac{1}{d_{i}d_{j}}$. Consequently, the self-information of the link, interpreted as its weight, is computed as: 

\begin{equation}
SI(i,j)  = - \log_2{P(i,j)} \\ 
         = \log_2({d_{i}d_{j}}).\nonumber
\end{equation}

For undirected links, the value of $SI(i,j) = SI(j, i)$. We now define two notations $S(i)$ and $S^+(i)$, which are important to calculate the node importance score. $S(i)$ denotes the sum of SI values of all links connected to node $i$ and is calculated by $S(i)  =  \sum_{j \in {\cal N}_{i}} SI(i,j)$. $S^+(i)$ captures the combined structural importance of links incident to node $i$ and its neighbors, defined as the sum of SI of links that node $i$ and its neighbors are one endpoint of these links, i.e., $S^+(i)  =  \sum_{j \in {\cal N}_{i} \cup \{{i}\} } S(j)$. Note that $S(i)$ accounts for the influence of links directly connected to node $i$, while $S^+(i)$ reflects a broader view by incorporating the link contributions of node $i$ and its neighbors. The probability assigned to any node $j \in {\cal N}_{i} \cup \{{i}\}$, denoted by $P(j)$, can be defined as 

\begin{equation}
P(j)  := \frac{S(j)}{S^+(i)},
\end{equation}

such that the sum of all the probabilities is equal to 1, i.e., $\sum_{j \in {\cal N}_{i} \cup \{{i}\}} \frac{S(j)}{S^+(i)} = 1$, and the IE of node $i$ is calculated as ${\cal E}(i) = - \sum_{j \in {\cal N}_{i} \cup \{{i}\} } P(j)\log_2{P(j)}$.

The amount of information a node possesses is influenced by the number and strength of its connections within the graph. To quantify this, we compute the expected entropy of node $i$, denoted by ${\cal E}(i)$, which incorporates the IE of neighboring nodes, weighted by link importance as a measure of relevance. Nodes are then ranked based on their expected entropy values: a higher ${\cal E}(i)$ indicates a more important node. In cases where multiple nodes share the same expected entropy, they are assigned the same rank. For example, in a network of six nodes, if nodes $1$ and $2$ have equal expected entropy, and nodes $4$ and $5$ share another common value, the ranking would be: ${\mathcal E}(1) = {\mathcal E}(2) > {\mathcal E}(3) > {\mathcal E}(4) = {\mathcal E}(5) > {\mathcal E}(6)$. To assess the importance of links within the graph, we transform the original graph into its corresponding line graph, where each node in the line graph represents an edge from the original graph \cite{de2020edge}. We then apply the IE-based method to this line graph. The resulting node importance vector in the line graph directly corresponds to the link importance in the original graph.

\subsection{Probabilistic Scheduling Design}
Similar to \cite{wang2022matcha, herrera2024faster}, we consider a probabilistic scheduling method, where an arbitrary number of collision-free subsets are activated in each communication round. The activation decision of a matching in each communication round of the D-SGD algorithm is an independent Bernoulli random variable $Z_j$, which is equal to 1 with probability $p_j$ and 0 otherwise, for each matching $G_j, \forall j \in \{1,...,M\}$. Here, $p_j$ represents the expected activation probability of the $j$-th matching.

Further, for node-based scheduling, each collision-free subset has a certain scheduling probability. Let the scheduling decision of collision-free subset $S_r$ in $k$-th communication round be $\textstyle{s_i}(k)$, where $\textstyle{s_i}=1$ if $S_r$ is activated, and $0$ otherwise. The graph of the communication topology activated in communication round $k$ is represented as $G(k) = \{V(k), E(k)\}$, where $G(k) = \bigcup_{i=1}^{q} \textstyle s_i(k) S_i$. Note that $G(k) = G^{\textit {d}}$ when all the collision-free subsets are activated in communication round $k$. The sum of the activation probabilities of all matchings (for link-based scheduling) and collision-free subsets (for node-based scheduling) should be less than or equal to the communication budget, denoted by $B$.

\subsection{Optimizing Scheduling Probabilities and Mixing Matrix}
We can obtain the topology of the scheduled broadcasting nodes by sampling the corresponding columns of the adjacency matrix of the base topology, leaving those columns unchanged, and masking the remaining columns as zero. The obtained matrix will represent a directed graph with edges connecting activated nodes (columns) to their neighbors. A network with bidirectional links can be obtained by sampling the columns and rows with the same indices (i.e., the $i$-th column and row) from the adjacency matrix. The resultant adjacency and Laplacian matrices will be symmetric, and as a result, the mixing matrix of the scheduled broadcasting nodes will also be symmetric.

Mathematically, a symmetric adjacency matrix $\mathbf{\hat {A}}$ can be obtained by sampling the columns and rows of $\mathbf{A}$, as $\mathbf{\hat {A}}(k) = \mathbf{Q}(k)\mathbf{A}\mathbf{Q}(k)$, where $\mathbf{Q}(k)= \diag(n_i(k),\ldots,n_N(k))$, with $n_i(k) = 1$ if node $i$ is activated in $k$-{th} iteration. It is important to note that scheduling decisions are made only on collision-free subsets, not on individual nodes, and multiple collision-free subsets can be scheduled in each communication round. The Laplacian matrix for the scheduled communication topology in $k$-{th} communication round is $\mathbf{\hat {L}}(k) = \diag (\mathbf{\hat {A}}(k)\mathbf{1}) - \mathbf{\hat {A}}(k)$, where the resulting Laplacian matrix $\mathbf{\hat {L}}(k)$ is symmetric, and the sum of each row and column is zero. Thus, the resultant mixing matrix for $k$-th communication round is also symmetric, defined as $ \mathbf{W}(k) = \mathbf{I} - \alpha \mathbf{\hat{L}}(k)$, where $\alpha$ is a constant independent of $k$.

\begin{figure*}[t]
    \centering
    \includegraphics[scale=1, height= 15 cm]{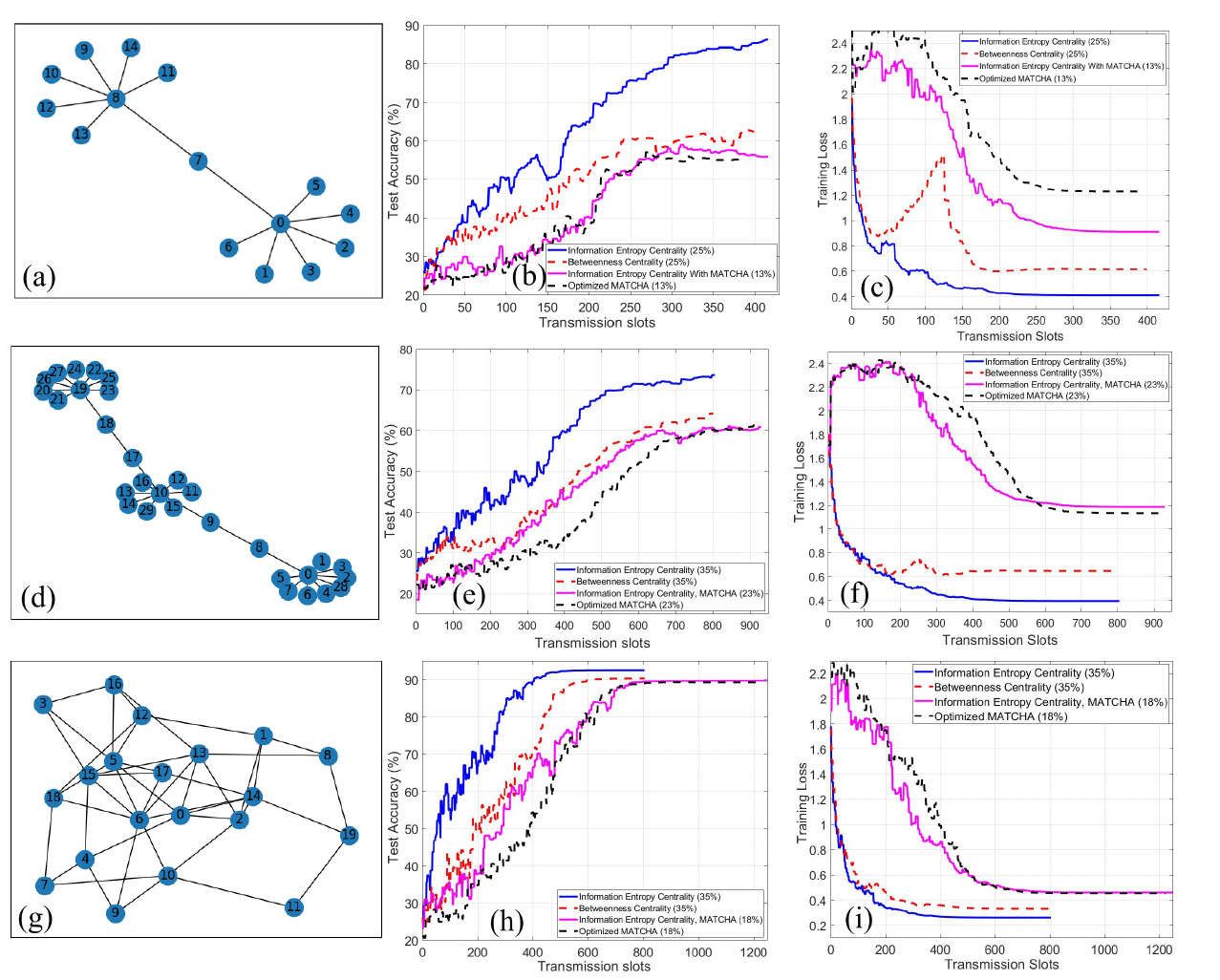}
    \caption{Performance comparison between IE and BC-based link/node importance methods.}
    \label{fig:results}
\end{figure*}

The scheduling probability of a networked node depends on its importance in the network topology, where an important node is scheduled more frequently than a less significant one. In every communication round, a subset $S_r$ is scheduled with probability $p_{S_r}$, for $ r \in \{1, \dots, q\} $, thus $\mathbb{E}[\mathbf{Q}(k)] = \diag (p_i, \dots,p_N)$.
Note that a node's activation probability is equal to the scheduling probability of the subset to which it belongs, i.e., $p_i = p_{S_r}$ if $i \in {\cal V}_r$. As a result, the number of subsets scheduled in a communication round is arbitrary, such that the average number of transmission slots per communication round is $\sum_{l=1}^q p_{S_l}$.

The subset scheduling probabilities had been optimized in \cite{herrera2024decentralized} to accelerate per-round convergence under constrained communication as follows:
\begin{equation}
\begin{aligned}
   \min_{p_{S_1}, \ldots, p_{S_q}} ||E(\mathbf{W}^2(k) - \mathbf{J})||_2, \\
  {\text{s.t.}} \quad \sum_{l=1}^q p_{S_l} = B \text{ and }
   0 \leq p_{S_l} \leq 1, \forall l.
   \label{eq:optimization}
\end{aligned}
\end{equation}

In \eqref{eq:optimization}, $\mathbf{J} = \frac{1}{N}\mathbf{11^T}$, where $\mathbf{1}$ is the all-ones column vector, and $B$ denotes the average count of transmission slots per communication round. For a detailed explanation of the optimization problem \eqref{eq:optimization}, the interested reader is referred to \cite{herrera2024decentralized}. In this work, we use the IE-based approach to rank networked nodes and compute their activation probabilities. Here, the activation probability of node $i$ is represented by $b_i$ such that $ \sum_{i=1}^N b_i = 1$, and the scheduling probability of a subset $j$ is given by $b_{S_j} = \sum_{i=1}^N b_i \mathbf{1}_{\{i \in {\cal V}_j} \},$ where the value of $1_{\{i \in {\cal V}_j} \}$ is equal to one if $i \in {\cal V}_j$, and zero otherwise. We also choose $p_{S_j} = \min \{1, \gamma b_{S_j}\}$, where the value of $\gamma$ is computed such that $\sum_{i=1}^q p_{S_i} = B$.

For every communication round, it is pivotal to design the mixing matrix as $\mathbf{W}(k) = \mathbf{I} - \alpha \mathbf{\hat{L}}(k)$ to enhance the convergence speed of D-SGD. Moreover, the mixing matrix can be optimized by solving (\ref{eq:optimization}) considering $\alpha$ as an optimization parameter. The solution to this convex problem is provided in \cite{herrera2024decentralized} and is reproduced here for the sake of completeness as follows:

\begin{equation}
\begin{aligned}
 &  \min_{s, \alpha, \beta} \quad s, \\
 &  \text{s.t.} \quad \alpha^2 - \beta \leq 0, \\
 &   \mathbf{I} - 2\alpha \mathbb{E}\left[\mathbf{\hat{L}}(k)\right] + \beta \left(\mathbb{E}\left[\mathbf{\hat{L}}^T(k)\mathbf{\hat{L}}(k)\right]\right) - \mathbf{J} \leq s\mathbf{I}
\end{aligned}
\end{equation}

where $s$ and $\beta$ are auxiliary variables.

\section{Experimental Results}
In this section, we evaluate the performance of the IE-based link- and node-importance metric in terms of test accuracy and training loss within the D-SGD framework, following the experimental setup described in \cite{herrera2024decentralized} that partitions the  MNIST dataset among nodes in non-IID fashion. Our experiments evaluate network topologies of up to 50 nodes, revealing consistent trends across 2-star, 3-star, random, and hierarchical topologies. However, due to space constraints, we report results for only three different network topologies comprising $N \in \{15, 20, 30\}$ nodes, as shown in Fig. \ref{fig:results}.
As illustrated in Fig. \ref{fig:results}, we compare the performance of IE-based and BC-based node importance metrics in three different network topologies. The percentage shown in each figure indicates the average fraction of activated subsets per iteration. While we experimented with various activation percentages across all topologies, we report results only for lower activation levels because of space limitations.
The IE-based node importance metric consistently achieves higher test accuracy in fewer transmission slots than the BC-based method when up to 60\% of subsets are activated in all three network topologies. For instance, in Fig. \ref{fig:results}(b), with 25\% subset activation in the 2-star topology, the IE-based approach reaches 70\% test accuracy in approximately 200 transmission slots, whereas the BC-based method fails to achieve 70\% accuracy even after 400 slots. 
A similar pattern is observed in the 3-star topology at 35\% subset activation: the IE-based approach achieves 70\% accuracy in roughly 500 slots, while the BC method remains below this threshold even after 800 slots, as shown in Fig. \ref{fig:results}(e). We observe similar trends for random topologies. At 35\% activation, the IE-based method reaches 90\% test accuracy in around 400 transmission slots, whereas the BC-based approach requires nearly 600 slots to attain the same accuracy, as illustrated in Fig. \ref{fig:results}(h).
Furthermore, the IE-based link importance metric achieves target accuracy using as few as, or fewer, transmission slots compared to the optimized MATCHA approach, as demonstrated in Figs. \ref{fig:results}(b), (e), and (h). For example, in the 3-star topology with 23\% activated matchings, the IE-based method attains 60\% test accuracy in roughly 650 transmission slots, while optimized MATCHA requires approximately 770 slots. Conversely, both methods reach about 90\% test accuracy at nearly 800 slots (Fig. \ref{fig:results}(e)).
Importantly, the training loss associated with the IE-based method remains consistently lower than that of both the BC approach and optimized MATCHA, aside from some initial fluctuations, as shown in Figs. \ref{fig:results}(c), (f), and (i). When the BC and IE-based importance metrics yield similar performance, the IE-based approach offers several key advantages. It is computationally efficient, relying solely on local node degree and one- or two-hop neighborhood information, which facilitates decentralized implementation with minimal communication overhead. The method adapts rapidly to topology changes through localized updates and employs entropy-based ranking to break ties effectively, ensuring robust node selection even in symmetric network structures. Overall, the proposed IE-based method outperforms the BC approach in large, dense, and sparse networks. Its advantage lies in capturing the richness of local information and the structural complexity of the network, enabling the identification of influential nodes based on their diverse neighborhood structures. This leads to significantly higher test accuracy with fewer transmission slots compared to the BC-based method.

\section{Conclusion}
In this work, we proposed an IE-based link and node importance metric to compute the scheduling probabilities of disjoint subsets of links or nodes, aiming to accelerate decentralized learning under partial communication constraints. The key novelty of the IE-based approach lies in its dual use of self-information-weighted links and neighborhood entropy to quantify node importance, embedding structural complexity into the ranking process. This makes the approach scalable and robust, outperforming traditional path- and degree-based metrics. Simulation results verify the effectiveness of our proposed method compared to existing link and node scheduling schemes across various network topologies, including dense random graphs and sparse, irregular structures. A promising direction for future research involves developing node-scheduling strategies that jointly consider the volume of data at each node and the distributions of local datasets.

\section*{Acknowledgements}
This work was supported in part by ELLIIT, Swedish Research Council, the French government under the France 2030 ANR program “PEPR Networks of the Future” (ref. 22-PEFT0010), the Huawei France-EURECOM Chair on Future Wireless Networks, the EU’s Horizon program Grant Agreement No 101139232 (SNS JU Project 6G-GOALS), and the European Research Council (ERC) under the European Union’s Horizon 2020 research and innovation programme (Grant agreement No. 101003431) 

\IEEEtriggeratref{5}
\bibliographystyle{IEEEtran}
\bibliography{refs}
\end{document}